\documentclass[%
 reprint,
 amsmath,amssymb,
 aps,
]{revtex4-2}

\usepackage{graphicx}
\usepackage{dcolumn}
\usepackage{bm}
\usepackage{color}

\begin{document}


\title{Chaotic  Bloch oscillations in dissipative optical systems driven by a periodic train of coherent pulses}

\author{A. Verbitskiy}
\affiliation{School of Physics and Engineering, ITMO University, Kronverksky Pr. 49, bldg. A, St. Petersburg, 197101, Russia}

\author{A. Balanov}
\affiliation{Department of Physics, Loughborough University, Loughborough, LE11 3TU, United Kingdom}

\author{A. Yulin}
\affiliation{School of Physics and Engineering, ITMO University, Kronverksky Pr. 49, bldg. A, St. Petersburg, 197101, Russia}

\date{\today}

\begin{abstract}
We study the response of an optical system with the Kerr nonlinearity demonstrating Bloch oscillations to a periodic train of coherent pulses. It has been found out that the intensity of the field excited in the system by  pulses resonantly depends on the train period. It is demonstrated numerically and analytically that the  response of the system is stronger when the period of the driving pulses is commensurate with the period of the Bloch oscillations. Moreover,  large enough pulses are  capable to induce the instabilities which eventually lead to onset of chaotic Bloch oscillations of the wave-function envelope bouncing both in time and space. The analysis reveals that these instabilities are associated with period-doubling bifurcations. A cascade of such bifurcations with increase of the pulses' amplitude triggers the chaotic behaviour.
\end{abstract}


\maketitle

\section{Introduction \label{sec:Introduction}}
Bloch oscillations is a very important fundamental phenomenon first discovered during the development of zone theory of the solid state physics \cite{discovery_1, discovery_2, discovery_3}. The effect manifests itself in a counter-intuitive periodic oscillations of quantum particles moving in a tilted periodic potential, e.g. electrons in a crystals subjected to a constant electric field. The theoretical discovery was followed by a long scientific discussion and finally the effect was confirmed experimentally \cite{discussion_1, discussion_2, discussion_3, discussion_4}, see also review \cite{discussion_review}.  

It is well known that, under some conditions, the dynamics of light is described by the same equations as the wave function of particles in quantum mechanics. Therefore, it would be reasonable to anticipate that an analogue of Bloch oscillations can be found in optical systems. Indeed, optical Bloch oscillations have been predicted in optical waveguide arrays and photonic crystals \cite{BO_optics1, BO_optics2, BO_optics3, BO_optics4, BO_optics5}. In that systems the effective refractive index depends linearly on a spatial coordinate and this plays a  role of a linearly growing part of the potential in quantum systems. The spectrum of the eigenmodes of these systems is equidistant (have a form of the Wannier-Stark ladder) and the light propagates along snaking trajectories. 

Optical Bloch oscillations are much easier to observe experimentally compared to their quantum counterparts and thus their theoretical discovery was accompanied by a number of experimental works where Bloch oscillations were demonstrated \cite{BO_optics1, BO_optics_exp1, BO_optics_exp2, BO_optics_exp3, BO_optics_exp4, BO_optics_exp5, BO_optics_exp6, BO_optics_exp7}. A comprehensive review on optical Bloch oscillations, Zener tunneling and related effects can be found in \cite{BO_optics_review}.  It is important to acknowledge that Bloch oscillations  are a very generic effect and occur in a large number of physical systems such as atomic systems \cite{atomic1, atomic2, atomic3, atomic4, atomic5, atomic6}, lasers \cite{laser}, coupled LC circuits \cite{LC}, mechanical systems \cite{mech1, mech2, mech3, mech4}, plasmonic \cite{plasmonic1, plasmonic2, plasmonic3, plasmonic4, plasmonic5, plasmonic6} or exciton-polariton systems \cite{exciton1, exciton2, exciton3}. 

Bloch oscillations is a linear phenomenon but, of course, nonlinearities of the physical systems can affect Bloch oscillations. In most cases the effect of the nonlinearity on the Bloch oscillations is destructive making it impossible to observe the oscillations at long times \cite{nonl_BO1, nonl_BO2, nonl_BO3, nonl_BO4, nonl_BO5, nonl_BO6}. The main reason why nonlinear effects prevent observation of long-living Bloch oscillations is modulation instability appearing when the envelope approaches the edges of the band \cite{nonl_BO_expl}. The understanding of this fact allowed to suggest different, sometimes quite complicated methods of nonlinearity management stabilizing Bloch oscillations in the nonlinear regime \cite{nonl_BO_stab1, nonl_BO_stab2, nonl_BO_stab3, nonl_BO_stab4}. It was also found that, quite surprisingly, the increase of the dimensionality of the system can also stabilize Bloch oscillations \cite{, nonl_BO_stab5} making possible, for instance, their use for resonant new frequency generation \cite{nonl_BO_rad}. On the other hand, it has been shown that nonlinear Schr\"{o}dinger equation are able to demonstrate deterministic chaos \cite{Chaos_NLS1, Chaos_NLS2, Chaos_NLS3, Chaos_NLS4}. Despite of the above progress, the non-linear response of the systems with Bloch oscillations to external pulse excitations is still poorly understood.

This paper aims to contribute to a better understanding of nonlinear regimes of optical Bloch oscillations in a system of interacting cavities pumped by \emph{a periodic train of external coherent pulses.} It is shown that the excitation efficiency greatly increases when the period of repetition of the pump pulses is commensurate with the period of Bloch oscillations. In the presence of coherent pump the dephasing of the Wannier-Stark states due to nonlinear effects can be safely neglected provided that the dephasing time is much longer compared to the lifetime of the waves excited in the system. Our analysis shows that the presence of the Kerr nonlinearity evokes symmetry breaking and promotes occurrence of the so-called period-doubling bifurcations when the intensity of the excited field exceeds some threshold. Such a bifurcation destabilises the Bloch oscillations with the given period and simultaneously gives birth to the Bloch oscillations with doubled period. A cascade of these bifurcation with further increase of excitations leads to appearance of deterministic chaos, which destroys any periodicity of the Bloch oscillations.  We study the correlation properties of the non-linear Bloch oscillations and also discuss the effects of dissipation on the development of chaos.

\begin{figure}
\includegraphics[width=0.45\textwidth]{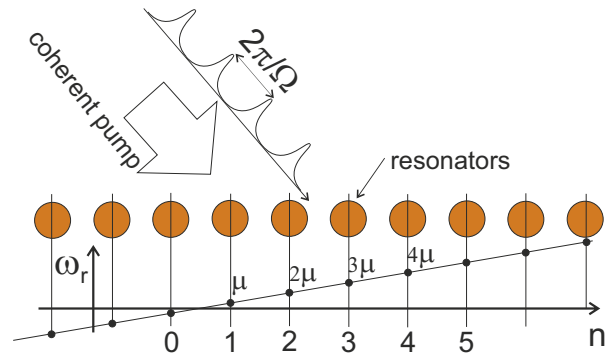}
\caption{(Color online) The schematic view  of the array of optical resonators pumped by a train of coherent pulses. The resonant frequencies of the resonators depend on their indexes linearly. }\label{fig0}
\end{figure}

In our work we consider a chaotic behaviour in a system consisting of a one-dimensional array of coupled single-mode resonators. The schematic view of the resonator array is given in Fig.~\ref{fig0}. The systems of such a kind, in particular based on the Bound State in the Continuum effect, are widely discussed now in the scientific literature in the context of second harmonic generation and other nonlinear phenomena \cite{BIC_resonators_Koshelev,BIC_resonators_Zograf,BIC_resonators_Jahani,BIC_resonators_Hwang}. Indeed, the Wannier-Stark states and Bloch oscillations in such systems are very promising for various sources of coherent radiation,  which relies on gain by resonances. The wide Wannier-Stark states increases the mode volume which gains the total power of the generated radiation.  In the same time the resonant frequencies of the states depend on their geometrical positions and this opens a possibility to tune the working frequency by just shifting the excitation spot. However, utilization of the above effects for practical applications requires understanding the dynamics of the pulses in driven-dissipative Bloch systems.

The resonators in Fig.~\ref{fig0} can be pumped by pulses of external coherent light of finite spatial aperture launched at some angle to the resonator array. The resonators are coupled to free propagating waves which means that the resonator modes are leaky and the resonators experience radiative losses. The total losses are the sum of the radiative and Joule losses and are characterized by an effective dissipation rate. To achieve Bloch oscillations in the system we make the resonant frequency of the resonators to be a linear function of the index numbering the resonators. We study the case where the resonant frequency of the resonators depends on the intensity of the field in the resonators and in the leading approximation the shift of the resonant frequency is proportional to the intensity of the mode. We assume the nonlinearity to be an instantaneous cubic one.  A simplest realization of such systems is an array of conventional optical nonlinear resonators. 

To study the influence of the losses on the dynamics of the system we consider a polaritonic system consisting of interacting micro-pillars pumped simultaneously by incoherent and coherent pumps. The incoherent pump is needed to control the effective losses seen by the polaritons. To study the nonlinear effects it is convenient to have high-Q resonators and this can be achieved by applying a proper incoherent pump. However, in this paper we keep the pump below the threshold where polariton lasing starts. The reason to consider a polariton system is that such systems exhibit very strong Kerr nonlinearity facilitating experimental investigation of the nonlinear effects. It is worth mentioning here that spatial period doubling has been already predicted and observed in polariton systems \cite{polariton_period_doubling}. Recently chaotic behaviour of polariton system driven by continuous radiation is also reported \cite{polariton_chaos}.

In contrast to previous works, here we analyse the regimes induced by trains of light pulses, i.e. by electromagnetic radiation with periodically varying intensity. In particular we focus on the effects of pulse frequency rather than frequency of electromagnetic wave. This allows to have high peak intensity of the field with relatively low average intensity and this way to decrease the heating of the sample. This can be of importance for nonlinear applications requiring high field intensities, for instance, the third harmonic generation.

The paper is structured as follows. In Section \ref{sec2} we consider a mathematical model capable to describe Bloch oscillations in driven-dissipative systems. In this section we discuss a linear regime of Bloch oscillations and show that the efficient excitation takes place when the field intensity is varying with the frequency equal to the frequency of Bloch oscillations. In Section \ref{sec3} the nonlinear propagation of the filed is considered. It is shown that the nonlinearity brakes the symmetry of the field propagating in the system. Chaotic behaviour of the system is also discussed in this section. Section \ref{sec4} is devoted to nonlinear dynamics of polariton systems. It is shown that period doubling bifurcation takes place in polariton systems too.  The main results of the paper are briefly summarized in the Conclusion.

\section{Linear regime of Bloch oscillations in an array of coherently driven resonators}
\label{sec2}

To model the system illustrated in Fig.~\ref{fig0}, we assume that the inter-resonator coupling, losses and the nonlinear effects do not change the structure of the field in each of the resonator, but affect the amplitudes and the phases of the resonator modes. Thus, a  tight-binding approximation can be used and the field in each of the resonator can be characterized by a slowly varying complex amplitude $u$. Then the dynamics of electromagnetic field in the system is described by a discrete dissipative nonlinear Schrodinger equation  written for the complex amplitudes of the resonator modes $u_n(t)$
\begin{widetext}
\begin{eqnarray}
i\partial_t u_n = \left( \mu n  -i \gamma  + \alpha |u_n|^2 \right) u_n -\sigma(u_{n+1}+u_{n-1}-2u_{n}) +a_n(t) \exp(-i\omega_p t+i k_p n), \label{nls1} 
\end{eqnarray}
\end{widetext}
where $n$ enumerates the resonators, $\gamma$ is linear losses, $\mu$ characterizes the steepness of the linear dependency of the eigenfrequency of the resonators  on their number, $\sigma$ is the coupling strength between the neighbouring resonators, $\alpha$ is the nonlinear coefficient, $a_n(t)$ is the amplitude of the pump coming to $n$-th resonator, $\omega_p$ is the detuning of the pump frequency from the resonance frequency of the resonator with $n=0$, $k_p$ is the projection of the phase gradient of the pump field on the axis passing through the resonators array. For the sake of mathematical convenience we introduce dimensionless units.   

\begin{figure}
\includegraphics[width=0.45\textwidth]{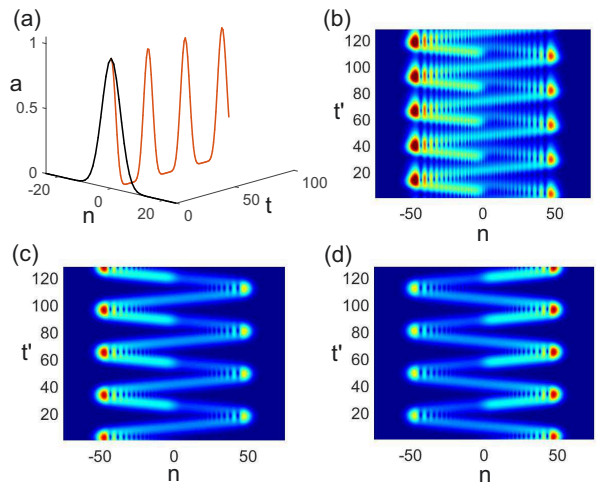}
\caption{(Color online) (a) the dependencies of the normalized driving force amplitude $a(n, t=0)$ and $a(n=0, t)$. The evolution of the field $u_n(t)$ in stationary regime are shown in panels (b) and (c) for the frequency of driving force amplitude $\Omega=0.24$ and $\Omega=0.2$ correspondingly. (d) is the same as (c) but for the opposite sign of the driving force wave vector, $k_p=-\pi/2$. The amplitude of the driving force is small to insure linear regime of propagation. The other parameters are $\gamma=0.025$, $\mu=0.2$.  }\label{fig1}
\end{figure}

We use the aperture of the excitation beam that is much smaller than the span of Bloch oscillations. At the same time, we require the aperture to be wide enough so that its spatial spectrum is narrow compared to the Brillouin zone. In our numerical simulation we took the aperture to be equal to $w=5$. Then considering the dynamics in the excitation spot we can neglect the dependency of the resonant frequency on the index of the resonator. In this case a single pulse excites a propagating envelope efficiently if the frequency and the wavevector of the pump are related as $\omega_p=2\sigma(1-\cos(k_p))+\mu n_p$ (the dispersion of the linear waves in the system), where $n_p$ is the position of the pump centre. Without loss of generality we pump at $n_p=0$. In the examples of numerical simulations presented in the paper the pump frequency is chosen to be in the middle of the zone, so $\omega_p=2\sigma$, $\sigma=5$, $k_p=\pi/2$. 
The variation of the amplitude of the pump exciting the system we take in the form $a_n(t)=a_p|\sin(\frac{\Omega}{2} t)|^7 \exp(-n^2/w^2)$ where  $\Omega$ is the frequency of the pump intensity variation.  The spatial distribution and the temporal evolution of the normalized driving force amplitude is shown in Fig.~\ref{fig1}.  

The linear regimes of propagation are illustrated in Fig.~\ref{fig1}. Each pulse of the pump excites a wave envelope experiencing Bloch oscillations. If the frequency of the pump pulses does not coincide with the frequency of Bloch oscillations then the waves excited by different individual pulses of the pulse train do not interfere constructively (panel (b) of Fig.~\ref{fig1}). The constructive interference occurs at $\Omega=m \mu$, where $m$ is an integer, see panel (c). The latter can be seen as a resonance between the pump and different Wannier-Stark states. Panel (d) of Fig. 2 illustrates the same case as in (c), but with $k_p=-\pi/2$. Comparison of (c) and (d) reveals the symmetry $k_p \rightarrow -k_p$.

\begin{figure}
\includegraphics[width=0.45\textwidth]{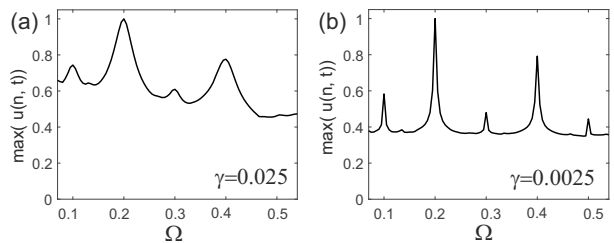}
\caption{The resonance curves showing the dependency of the maximum amplitude of the stationary field on the frequency of the driving force amplitude $\Omega$ for relative large losses $\gamma=0.025$ (a) and for smaller losses $\gamma=0.0025$. The amplitude of the driving force is very low so the the problem is linear.}\label{fig2}
\end{figure}

To characterize the Bloch resonance we calculated the maximum amplitude of the stationary field as a function of the frequency of the driving force amplitude. The results are presented in Fig.~\ref{fig2} showing the resonance curve for different losses. The maxima corresponding to the resonances are well developed even for relatively large losses.     

It is worth mentioning that the maxima of the response are also observed at the pump frequencies $\Omega=\tilde m \mu /m$, where $\tilde{m}$ is another integer. From the physical point of view this can be treated as a resonance of $m$-th harmonic of the pump with a Wannier-Stark state having resonant frequency $\tilde m \mu$. A simple mathematical consideration explaining the maxima at the frequencies $\Omega=\tilde m \mu /m$ are given in the Appendix. Let us remark that this special structure of the resonances can significantly facilitate the chaotization of Bloch oscillations.

It should also be noted that in the linear regime there is no difference between the excitation of the pulses propagating from the right to the left and the pulses propagating from the left to the right, compare Fig. 2(c) and (d) corresponding to pump wavevectors of the opposite signs. Another important remark to be made is that the oscillations are well localized in the linear regime.

\section{Nonlinear evolution of the field of driven Bloch oscillations}
\label{sec3}

\begin{figure} [h]
\includegraphics[width=0.45\textwidth]{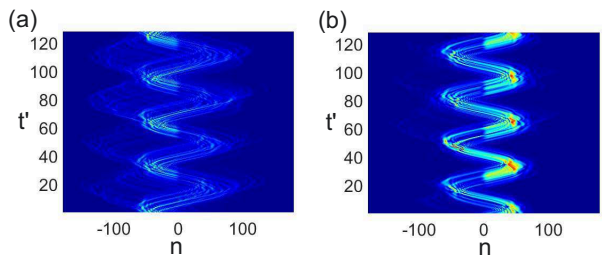}
\caption{(Color online) The stationary evolutions of the field in strongly nonlinear regime $a_p=0.8$ for $k_p=\pi/2$ (a) and $k_p=-\pi/2$ (b) for  $\Omega=0.2$ and $\mu=0.2$.} 
\label{fig3}
\end{figure}

Now let us consider how Kerr effect (linear dependency of the resonant frequencies of the resonators on the field intensity) affects the dynamics of the system. The first important observation is that in the nonlinear regime the symmetry $k_p \rightarrow -k_p$ is broken. To illustrate this symmetry breaking we take a very intense pump with the amplitude $a_p=0.8$ and calculate the dynamics for the wavevectors of the opposite signs $k_p=\pm\pi/2$. In Fig.~\ref{fig3} the  stationary stages of field propagation are shown for these parameters. It is clearly seen that the dynamics is very different for different signs of $k_p$. For lower pump intensity the asymmetry also takes place but is not that pronounced.  

It is also worth mentioning that in the strong nonlinear regime the localization of the field become larger. The spreading of the field cannot be explained by the inter-band tunneling because the dispersion has only one branch. The generation of the new waves gives rise to the formation of numerous snakes of Bloch oscillations. 

\begin{figure}
\includegraphics[width=0.45\textwidth]{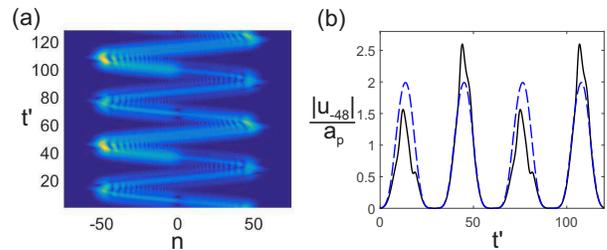}
\caption{(Color online) (a) The stationary evolution of the field for the pump amplitude $a_p=0.3$. The temporal evolution of the absolute value of the field at the site $n=-48$ is shown in panel (b) by solid black line. The dashed blue line shows the evolution of the field amplitude at the same site for the pump amplitude $a_p=0.2$. For convenience both amplitudes are normalized on the amplitude of the pulse $a_p$ for  $\Omega=0.2$ and $\mu=0.2$.}
\label{fig4}
\end{figure}

Let us now study in more detail what happens to the dynamics of the field when the driving force amplitude increases. The evolution of the field excited by the driving force of the amplitude $a_p=0.3$ is shown in panel (a) of Fig.~\ref{fig4}. It is seen that the period of the temporal oscillations became twice of that in the linear regime. To make it even more obvious we plotted the dependency of the amplitude of the field at site $n=-48$ as a function of time for the pump $a_p=0.3$, see panel (b) of Fig.~\ref{fig4}. For reference we plotted the same dependence but for the pump $a_p=0.2$ by the blue line. The pump $a_p=0.2$ is already nonlinear regime, but the period is the same as in the linear regime. This gives a reason to suggest that the period doubling bifurcation takes place in the system. Below we prove that this indeed what happens in the system and that a chain of period doublings results in the chaotic behaviour of the system.

\begin{figure}
\includegraphics[width=0.45\textwidth]{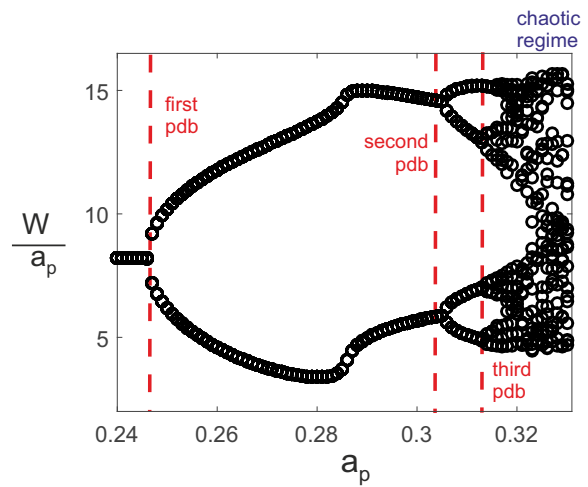}
\caption{(Color online) The figure shows the Feigenbaum diagram calculated for the resonantly pumped Bloch oscillations in the system of coupled oscillators described by (\ref{nls1}). The horizontal axis is the pump, the vertical axis is the energy $W=\sum_n |u_n|^2$ normalized on the pump amplitude $a_p$. The energy $W$ is measured at the points where the phase trajectory crosses the hyper-surface defined by the condition $t=2m\pi/\Omega$. The positions of the first three period doubling bifurcations (pdb) are marked by the dashed red vertical lines and labeled with "first pdb", "second pdb" and "third pdb" correspondingly. } \label{fig9-1}
\end{figure}

For our calculations, we set the values $\Omega=0.2$ and $\mu=0.2$, which in the case of linear response correspond to the principal resonance $1/1$, and analyse how the stationary regimes change with variation of $a_p$. The discussed effect is most pronounced if the system is excited at the main resonance and this motivated our choice of the pulse sequence period. The comprehensive studies of the off-resonant excitation is of interest but is out of the scope of the present paper. 

The map of regimes is summarised by the bifurcation diagram in Fig. \ref{fig9-1}. The figure illustrates the stroboscopic section of energy $W=\sum_n |u_n|^2$ whose values were taken in the discrete time moments $t=2m\pi/\Omega$, $m=1,2,\ldots$. For a particular value of $a_p$, the periodic oscillations are represented by one or few single points, and many points for the same $a_p$ correspond to chaos. The vertical dashed lines indicate the values of $a_p$ above which the number of points doubles, i.e attribute the period-doubling bifurcations. For small $a_p$ the system demonstrates period-1 oscillations, which are represented by a single point for the given value of $a_p$. At $a_p\approx 0.245$, the period of oscillations doubles, which is manifested in the appearance of pair of points on the diagram. The next period-doubling bifurcation giving rise to the period-4 oscillations takes place at $a_p\approx 0.306$. An infinite cascade of such bifurcation leads to emergence of the deterministic chaos, which exists for $a_p>0.318$. 

\begin{figure}
\includegraphics[width=0.45\textwidth]{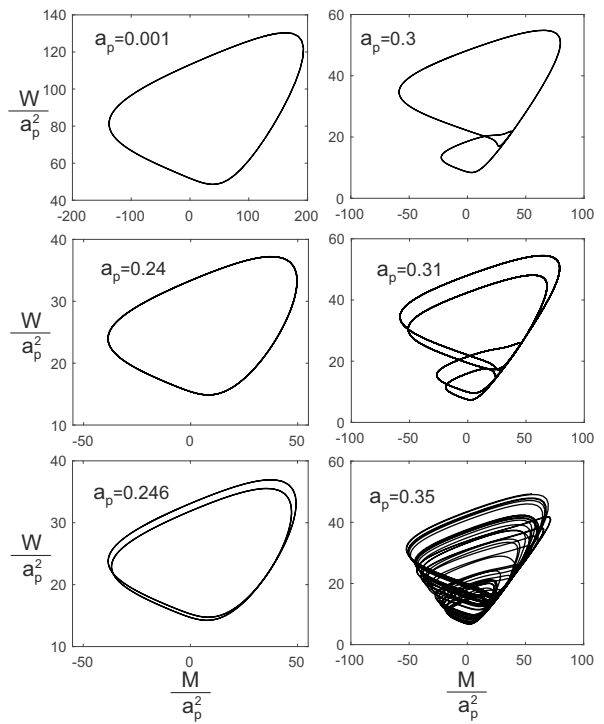}
\caption{The projections of the phase trajectories onto the phase plane ($W=\sum_n |u_n|^2$;$M=Im \left( \sum u_n^{*}\cdot ( u_{n-2}-u_{n-1})  \right)$) for different amplitudes of the resonant pump $a_p$. The other parameters are the same as in Fig.~\ref{fig4}.}
\label{fig5}
\end{figure}

To illustrate different dynamical regimes and the transitions between them, in Fig.~\ref{fig5} we plot the projection of the phase trajectories calculated for different $a_p$ on the space plane $(M, W)$, where $M=Im \left( \sum u_n^{*}\cdot ( u_{n-2}-u_{n-1})  \right)$ is a momentum. These variables are convenient for illustration since in the linear regime they both are proportional to $a_p^2$, which simplifies comparison.

In close to linear regime  the shape of the trajectory does not change qualitatively but its span decreases, see the trajectory calculated for $a_p=0.24$. However, the shape of the trajectory dramatically changes when the pump amplitude exceeds a certain threshold value. Then the trajectory makes a twist, and now it can be considered as consisting of two similar loops, see the trajectory calculated for $a_p=0.246$. This is a characteristic feature of period doubling bifurcation. 

At higher pumps the trajectory continues to deform, see the one calculated for $a_p=0.3$. Let us remark that between $a_p=0.246$ and $a_p=0.3$ no period doubling bifurcation happens, in this pumps interval the trajectory transform smoothly, but quite significantly. That is why we show the case just after period-2 bifurcation ($a_p=0.246$) and just before period-4 bifurcation ($a_p=0.3$). Then another period doubling bifurcation occurs, see the trajectory calculated for $a_p=0.31$. Finally, these period doubling bifurcations result in very complex behaviour, see the trajectory calculated for $a_p=0.35$.    

\begin{figure}
\includegraphics[width=0.45\textwidth]{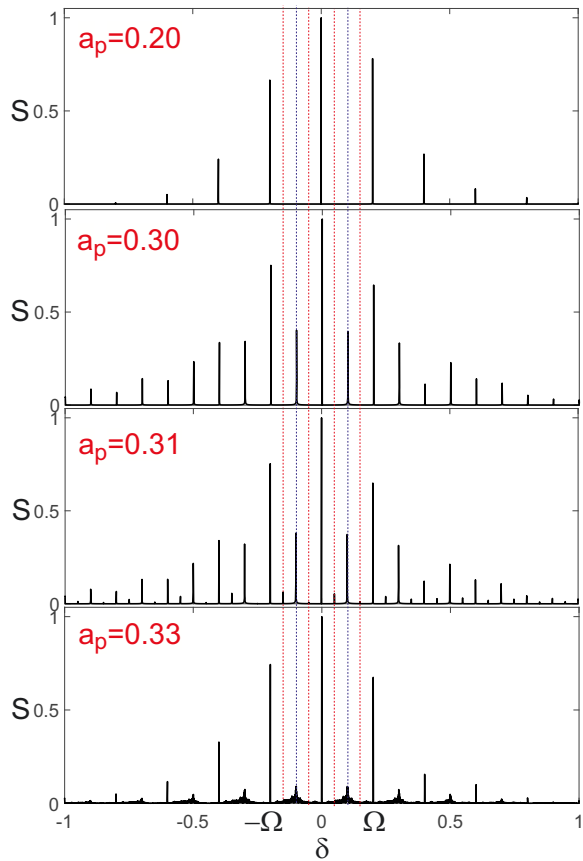}
\caption{(Color online) Temporal spectra of the field at site $n=-48$ for different amplitudes of the driving force $a_p$.  Dashed blue and red lines show the positions of several first subharmonics at $l\Omega/2$ and $l\Omega/4$. The horizontal axis is $\delta=\omega-\omega_p$ the detuning of the frequency from the carrier frequency of the pump. }\label{fig6}
\end{figure}

Evolution of oscillation spectra on the way to chaos are illustrated in Fig.~\ref{fig6}. We took a stationary variation of the field at site $n=-48$ and calculated its spectra for different levels of the pump. For weak excitations $a_p$, the spectrum contains the carrier frequency and the harmonics detuned from the carrier frequency by the frequency of the oscillations of the pump amplitude. After the period doubling the subharmonics appear. These subharmonics at the frequencies $l \Omega/2$, where $l$ is an integer, indicate that the period of the oscillations becomes twice of the period of the driving force. Then the next period doubling bifurcation produces the subharmonics at the frequencies $l \Omega/4$. Finally, we arrive to the spectrum of the chaotic signal consisting of two parts: the continuous background overlapped with the discrete spectrum. The first one corresponds to the chaotic part of the field and the second one - to the regular component of the field that can be seen as direct response of the system to the driving force. Let as note that, of course, the number of harmonics defines the shape of the field, but not its period.

\begin{figure}[t]
\includegraphics[width=0.5\textwidth]{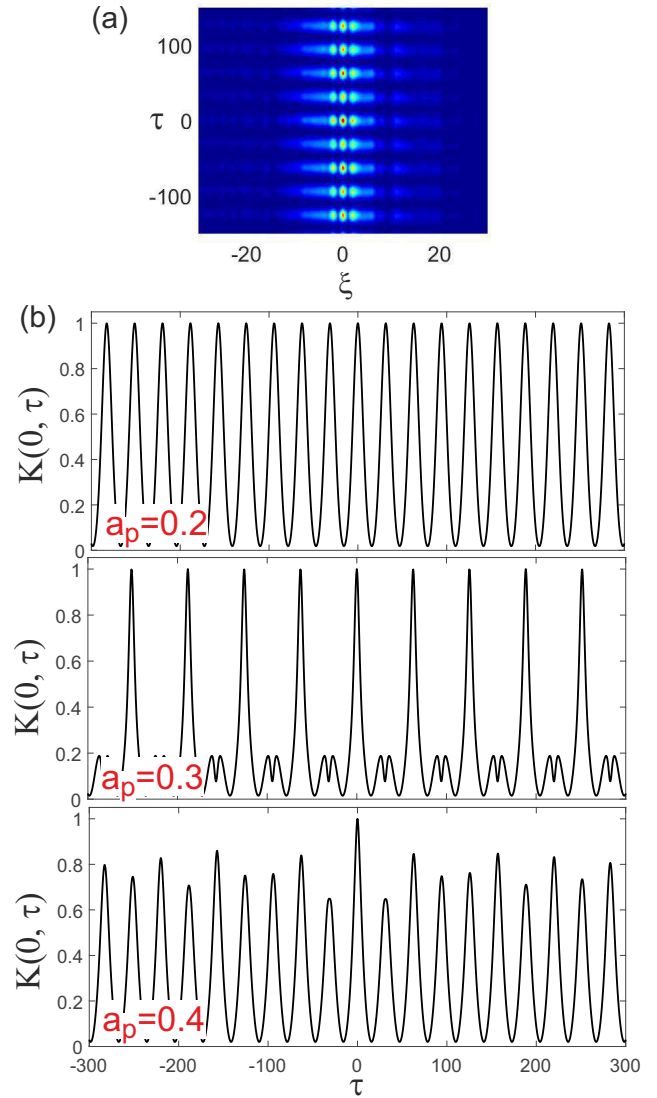}
\caption{(Color online) (a) Two-dimensional correlation function calculated for the stationary field for driving force amplitude $a_p=0.4$. The sections of the two-dimensional correlation functions $K(\xi=0, \tau)$ for different pump amplitudes are shown in panel (b).}   \label{fig7}
\end{figure}

To study the correlation properties of different regimes in the system we calculate the correlation function of the field defined as
$$ K(\tau, \xi)=\lim_{T\rightarrow \infty  N\rightarrow \infty} \frac{1}{4NT} \sum_{n=-N}^{N} \int_{-T}^{T} u_{n-\xi}(t-\tau)\cdot u_{n}(t) dt.  $$ 
The correlation function is calculated by the averaging over a large window where the signal is stationary. In numerics the function depends on the position of the averaging window, but we have checked that this difference is small.  The correlation functions are shown in Fig.~\ref{fig7} for $a_p=0.4$, $N=256$, $T=1000$.  

It is seen that the correlation function is well localized along the discrete coordinate, this is so because the field is well localized along the discrete coordinate. It is more interesting how the correlation function changes with time. The dependencies of the correlations on time are shown in panel (b) of Fig. 9 for different regimes of the field propagation. It is seen that for low pumps the correlation function is periodic. After the period doubling the period of the correlation function also becomes doubled. In the chaotic regime the correlation function rapidly drops down to some finite background value implied by application of a periodic driving. Thus, we can conclude that in the presence of cubic nonlinearity the resonantly excited Bloch oscillations of the coherent light can switch to chaotic regime via period doubling bifurcation. 

Next, we study how dissipation (including the nonlinear one) affects this phenomenon. For this purpose we consider a polariton system that possesses high nonlinearity due to strong light matter coupling and can be seen as a promising system where the discussed effects may be observed.  

\section{Polariton systems}
\label{sec4}
One of the important examples of the dissipative non-liner systems capable of demonstrating Bloch oscillations is coupled semiconductor microcavities (pillars) supporting exciton-polaritons \cite{exciton1,exciton2,exciton3}. The systems of this kind are promising for experimental verification of the effect because they are highly nonlinear and allow to control effective linear losses making them small enough. To describe the polariton dynamics we use the widely accepted model \cite{polariton_model} consisting of the equation for the order parameter of the polariton field $\psi$ and the density of incoherent excitons $\rho$
\begin{widetext}
\begin{eqnarray}
i\partial_t \psi_n = \left( \mu n  -i \gamma_1 + i \rho_n +|\psi_n|^2 +\alpha \rho_n \right) \psi_n- \sigma(\psi_{n+1}+\psi_{n-1}-2\psi_{n})+
Q_n \exp(i\omega_p t-i k_p n), \label{m_eq1} \\
\partial_t \rho_n = -\left( \gamma_2 +\beta |\psi_n|^2 \right)\rho_n + P_n. \label{m_eq2}
\end{eqnarray}
\end{widetext}
where $\sigma$ accounts for the discrete diffraction, $\gamma_1$ is the losses for the coherent polaritons, $\alpha$ accounts for the blue shift of the coherent polaritons due to their nonlinear interaction with the incoherent ones, $\gamma_2$ is the losses in the incoherent polaritons subsystem, $\beta$ defines the additional damping rate of the incoherent polaritons caused by their condensation into the coherent polaritons, $P_n$ is the amplitude of the incoherent pump, $\mu$ is the gradient of the resonant frequency, $Q_n(t)$ is the amplitude of the coherent pump, $\omega_p$ is its frequency and $k_p$ is its wave vector.
We use dimensionless units normalizing the time by the typical for such systems characteristic time $10$ ps; the polariton density is normalized by the characteristic polariton density $\frac{\hbar }{g_c\tau}$ where $g_c$ is characteristic polariton-polariton interaction constant; the density of the incoherent exciton reservoir is normalized on $\frac{2}{R\tau}$  where $R$ is the condensation rate; $\alpha=\frac{2 g_r}{\hbar R}$ where $g_r$ is the coefficient characterizing the blue shift of the polariton frequency proportional to the incoherent excitons density; $\gamma_2=\Gamma_x \tau$ where $\Gamma_x$ is the relaxation rate of the reservoir; $\beta=\frac{\hbar R}{g_c}$. The polariton parameters vary for different experimental realizations and we take typical polariton parameters $g_c=6\cdot 10^{-3}$  meV$\cdot\mu$m$^2$,  $R=0.005$ ps$^{-1}\cdot\mu$m$^2$,  $g_r=1.2\cdot 10^{-2}$ meV$\cdot\mu$m$^2$,   $\Gamma_x=3$ ps$^{-1}$  \cite{exciton1,polariton_param}.

We performed numerical simulations for a realistic polariton decay rate $\tau=25$ ps, so the dimensionless constants are $\gamma_1=0.4$, $\alpha=7.3$, $\gamma_2=3.3$, $\beta=0.55$. The dimensionless coupling strength between polariton pillars is chosen to be $\sigma=5$ ($0.5$ ps$^{-1}$ in physical units), the difference in resonant frequencies in the neighbouring pillars is $0.2$ ($0.02$ ps$^{-1}$ in physical units). Let us remark that the polariton losses are too high for convenient observation of Bloch oscillation, but the losses can be compensated by the the reservoir created by the incoherent pump. Therefore we consider the case where the incoherent excitons density is close to the condensation (polariton lasing) threshold (the threshold is $\rho_{th}=\gamma_1$ in dimensionless units).

\begin{figure}
\includegraphics[width=0.45\textwidth]{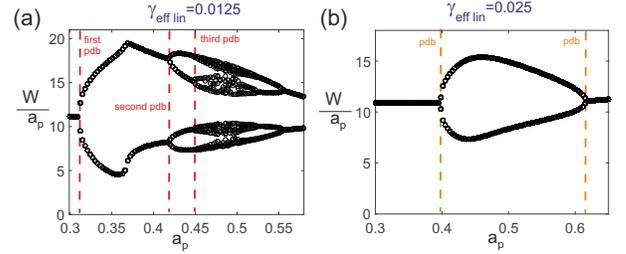}
\caption{(Color online) The figure shows Feigenbaum diagrams for the exciton-polariton system simultaneously pumped by the coherent and incoherent light. The incoherent pump is below but close to lasing threshold so that linear polaritons see effective losses $\gamma_{eff \, lin}=0.0125$ for panel (a) and $\gamma_{eff \, lin}=0.025$ for panel (b). The horizontal axis is the coherent pump amplitude $a_p$, the vertical axis is the energy of the field $W=\sum_n |\psi_n|^2$ divided by the pump amplitude $a_p$ calculated at the point where the phase trajectory crosses the hyper-surface defined by the condition $t=2\pi m/\Omega$. The positions of the period doubling bifurcations (pdb) are marked by the dashed red (in panel (a)) and orange (in panel (b)) vertical lines.} 
\label{fig10}
\end{figure}

For our numerical modeling we take the coherent pump in the same form as before $Q_n=a_p|\sin(\frac{\Omega}{2} t)|^7 \exp(-n^2/w^2)$. We calculated Feigenbaum diagrams for different incoherent pumps controlling the effective losses seen by the polaritons. Two examples are shown in Fig.~\ref{fig10} for the incoherent pump slightly below the lasing threshold so that the effective linear losses seen be the polaritons are low being equal to  $\gamma_{eff \, lin}=0.0125$ for panel (a) and $\gamma_{eff \, lin}=0.025$ for panel (b). The effective losses are defined as the losses seen by polaritons in the linear regime in the presence of the incoherent pump $\gamma_{eff \, lin}=\gamma_1-\rho_0$ where $\rho_0=P/\gamma_2$ is the exciton reservoir density created by the spatially uniform incoherent pump $P_n=P$ in the absence of the polaritons.

The carrier frequency of the coherent pump is chosen to be in resonance with the linear polaritons and in the presence of the reservoir of density $\rho_0$. Let us note that this resonant frequency depends on the intensity of incoherent pump $\omega_p=2\sigma+ \alpha \rho_0$ for $k_p=-\pi/2$ and thus the carrier frequency of the pump is slightly different for Fig. 10(a), (b). The frequency of the pulses sequence $\Omega$ is chosen to be in the main resonance with the Bloch oscillations $\Omega=\mu=0.2$. It is seen that for polariton systems the period doubling bifurcation occurs and in this sense the polariton systems are similar to the systems considered above.

The important difference, however, is that in the polariton system the chaotic regime can be achieved only if the incoherent pump is extremely close to the lasing threshold, which means very low linear losses. For the higher losses, as it is seen in Fig.~\ref{fig10}(b), the increase of the pump first evokes a period doubling of the Bloch oscillations at $a_p \approx 0.395$, but with further increase of the pump the period-1 Bloch oscillations (with period of the driving force) become stable again at $a_p \approx 0.618$.

One of the reason of such a behaviour of the system is that the origin of Bloch oscillations stochastization is the wave envelopes excited by different pulses of the pump interact to each other nonlinearly. The losses decrease the intensities of the interacting envelope making the nonlinear interaction less efficient. Polartions experience both the linear and nonlinear losses. The nonlinear losses also contribute to the decrease of the interaction efficiency of the different wave envelopes. Thus the nonlinear losses of the polariton make it more difficult to observe a full chain of period doubling bifurcations leading to the stochastic dynamics.  

\begin{figure}
\includegraphics[width=0.45\textwidth]{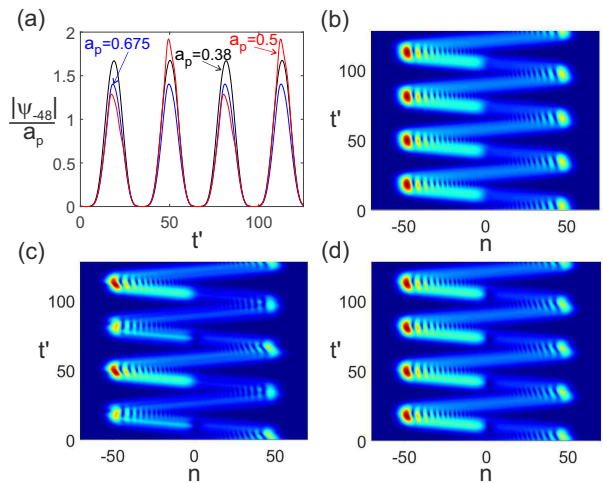}
\caption{(Color online) The dependencies of the normalized amplitudes of the field in site $n=-48$ on time are shown in panel (a) for the polaritons driven by the resonant coherent pump with the amplitudes $a_p=0.38$ (black line), $a_p=0.5$ (red line) and $a_p=0.675$ (blue line). Panel (b), (c) and (d) show spatial-temporal evolution of the filed for $a_p=0.38$, $a_p=0.5$  and $a_p=0.675$   correspondingly. The incoherent pump is chosen to provide effective losses $\gamma_{eff \, lin}=0.025$ in the linear regime of polaritons propagation.} \label{fig11}
\end{figure}

The dynamics of the field amplitude is illustrated in Fig.~\ref{fig11} showing the evolution of the field for the pump amplitudes $a_p=0.38$, $a_p=0.5$ and $a_p=0.675$. It is seen that for $a_p=0.38$ all Bloch oscillations are identical whereas for  $a_p=0.5$ the odd and the even Bloch oscillations become different. For the even higher pump  $a_p=0.675$  Bloch oscillations are identical again. Comparing panels (b) and (d) one can conclude that the dynamics of the fields excited by pumps $a_p=0.38$ and $a_p=0.675$ are very similar. 

We would like to acknowledge that the resonant excitation at fractional resonance $p/q$ can also be of interest as well as non-resonant excitation. The comprehensive studies of these cases are out of the scope of this paper and will be done somewhere else. Here we note that the excitation at the main resonance allows to achieve higher field intensities and thus to facilitate the observation of the nonlinear effects. This explains why this resonance case of greater importance.

\section{Conclusion}
\label{sec5}
In this paper we consider Bloch oscillations in the nonlinear driven-dissipative systems excited by a periodic train of coherent pulses. It is shown that in the linear regime the evolution of the field does not depend on the sign of the wavevector of the driving force. The efficiency of the excitation of the Bloch oscillations depends not only on the frequency of the field (temporal derivative of the phase of the field at a fixed site), but also on the period of the sequence of the pulses pumping the system. The maximum efficiency is achieved when the delay between the pulses is equal to the inverse Bloch frequency multiplied by $2\pi$.

The nonlinearity brakes the symmetry in the sense that the pulses launched in one direction propagate differently than the pulses launched in the opposite direction. More importantly, the nonlinearity causes period doubling bifurcation and the sequence of these bifurcations makes the Bloch oscillations to be chaotic. The field evolution can still be seen as Bloch oscillations, but every round of the oscillations is characterized by a different intensity of the field. It is also shown in the paper that coherently driven Bloch oscillations can be observed in exciton-polariton systems with experimentally achievable parameters. The period doubling bifurcation can occur in this system. 

Thus we can summarize that resonantly excited Bloch oscillations may be observed in nonlinear optical systems, including polariton ones, for the parameters realistic from the experimental point of view. The systems can demonstrate a complex dynamics resulting in new frequency generation and, under certain conditions, in the chaotization of Bloch oscillations through a chain of period doubling bifurcations.

\section*{Acknowledgements}

AV and AY acknowledge financial support from Priority 2030 Federal Academic Leadership Program and from Goszadanie no. 2019-1246 of the Ministry of Science and Higher Education of Russian Federation. 

\section*{Appendix}

Let us consider the dynamics described by Eq.(\ref{nls1}) in the linear limit $\alpha=0$. For sake of convenience we represent the amplitude of the driving force as $a_n(t)=\xi(n) A(t)\exp(-i\omega_p t)$. In the simulations shown in the main parts of the paper we took $A=a_p|\sin ( \frac{\Omega}{2} t)|^7 $
and $\xi=\exp (- n^2/ w^2) \exp(i k_p n)$, however here we consider a more general case $A$ being a periodic function of time. 

We notice that $W_m(n)=J_n(\frac{2\sigma}{\mu})$ are the eigenfunctions of the system so that $\mu n W_m(n)+\sigma(2W_m(n)-W_m(n+1)-W_m(n-1))=(2\sigma+\mu m)$, see \cite{BO_optics2}. The functions $W_m$ are orthogonal $\sum_n W_{m}(n) W_{m'}(n)=\delta_{m \, m'} $ ($\delta_{m \, m'}$ is Kronecker symbol) and so it is convenient to look for a solution of (\ref{nls1}) in the form
\begin{eqnarray}
u_n(t)=\sum_m C_m(t) \exp(-i\omega_p t)  W_m(n)  \label{eq_app_1} 
\end{eqnarray}
 Thus we obtain the equations for the coefficients $C_m$
\begin{widetext}
\begin{eqnarray}
i\partial_t C_m = (2\sigma +\mu m - \omega_p )C_m -i\gamma C_m+ A(t) \sum_n W_m(n) \xi(n). \label{eq_app_2} 
\end{eqnarray}
\end{widetext}

Since $A$ is a periodic function of time the stationary solution of (\ref{eq_app_2}) can be sought as Fourier series $C_m=\sum_l C_{ml} \exp(-i l\Omega t)$ where $\Omega=\frac{2\pi}{T_0}$ and $T_0$ is the period of the driving force amplitude. The expressions for $C_{ml}$ are easy to obtain
\begin{eqnarray}
C_{ml}(\Omega)=\frac{{\cal A}_l (\Omega) \sum_n \xi(n)W_m(n)}{ \omega_p +l\Omega -2\sigma -\mu m +i\gamma   }, \label{eq_app_3} 
\end{eqnarray}
where ${ \cal A}_l  = \frac{1}{T_0} \int_{-T_0/2}|^{T_0/2} A(t) \exp(i l \Omega t)dt$ are the Fourier representation of the amplitude $A$.

One can see that the for some frequencies the denominator reaches its minimum and this makes the dependency of $|C_{ml}(\Omega)|^2$ to be resonant. The resonances can be pronounced provides that the losses $\gamma$ are small and ${\cal A}_l(\Omega)$ is a flat function of $\Omega$. It is important to notice that the resonance appears when $\omega_p+l \Omega-2\sigma -\mu m=0$. This means that resonant frequencies are $\Omega_r=\mu \frac{m}{l}+ \frac{2\sigma-\omega_p}{l}$. In the section II the simulations we performed for $\omega_p=2\sigma$ and so one could expect to see the increase of the field intensity when the frequency of the amplitude modulation is equal to $\Omega=\frac{m}{l}$. These maxima are seen well in Fig.\ref{fig2}. 

To shed more light on the origin of these resonances let us consider the case of special driving force. We choose the spatial distribution of the driving force in the form of a Wannier function $\xi=W_{m0}(n)$. Then all $C_{m \neq m0 \, l}$ are equal to zero. Physically it means that the pump shaped as a Wannier function excites only one eigenmode of the system. The nonzero coefficients $C_{ml}$ are given by  
\begin{eqnarray}
C_{m0 \, l}(\Omega)=\frac{{\cal A}_l (\Omega) }{ \omega_p +l\Omega -2\sigma -\mu m_0 +i\gamma   } \label{eq_app_4} 
\end{eqnarray}
and the total field $u$ is expressed as
\begin{eqnarray}
u_n(t)=   W_{m0}(n) \exp(-i\omega_p t) \sum_l C_{m0 \, l}   \exp(-i l\Omega t). \label{eq_app_5} 
\end{eqnarray}

To characterize the efficiency of the excitation it is convenient to introduce such quantity as the total energy of the field averaged over a period of the driving force variation $E=\sum_n \frac{1}{T_0}\int_{0}^{T_0} |u_n|^2 dt$. Substituting here the field $u_n(t)$ in the form (\ref{eq_app_5}) we obtain a simple expression for $E$ through the coefficients $C_{m0 \, l}$
\begin{widetext}
\begin{eqnarray}
E_{m0}=   \sum_l |C_{m0 \, l}|^2=\sum_l \frac{|{\cal A}_l|^2 }{(\omega_p +l\Omega -2\sigma -\mu m_0)^2+\gamma^2}. \label{eq_app_6} 
\end{eqnarray}
\end{widetext}

Let us analyze (\ref{eq_app_6}) for the pump with $\omega_p=2\sigma$. The average energy $E$ does not show any resonances if pump excites the eigenmode with spatial distribution in the form of the Wannier function $W_{0}(n)$. If the pump is coupled to the mode $m_0=1$ then the dependency (\ref{eq_app_6}) has resonance maxima at the frequencies $\Omega=\frac{\mu}{l}$. Excitation of the mode with $m_0$ gives the resonances at $\Omega=\frac{m_0 \mu}{l}$. 

It is good to notice that the expression (\ref{eq_app_6}) can be simplified for the case of pump in the form of a train of the very narrow pulses. Then the dependency of ${\cal A}$ on $l$ is flat and calculating the sum in (\ref{eq_app_6}) we can neglect this dependency assuming ${\cal A}_l \approx {\cal A}_0$. Then the sum can be calculated analytically 
\begin{eqnarray}
E_{m0}=   \frac{|{\cal A}_0|^2}{\Omega} \frac{\sinh(\frac{2\pi\gamma}{\Omega})}{\cosh(\frac{2\pi\gamma}{\Omega})-\cos(\frac{2\pi(\omega_p-2\sigma - m_0 \mu)}{\Omega})}. \label{eq_app_7} 
\end{eqnarray}

\begin{figure}
\includegraphics[width=0.45\textwidth]{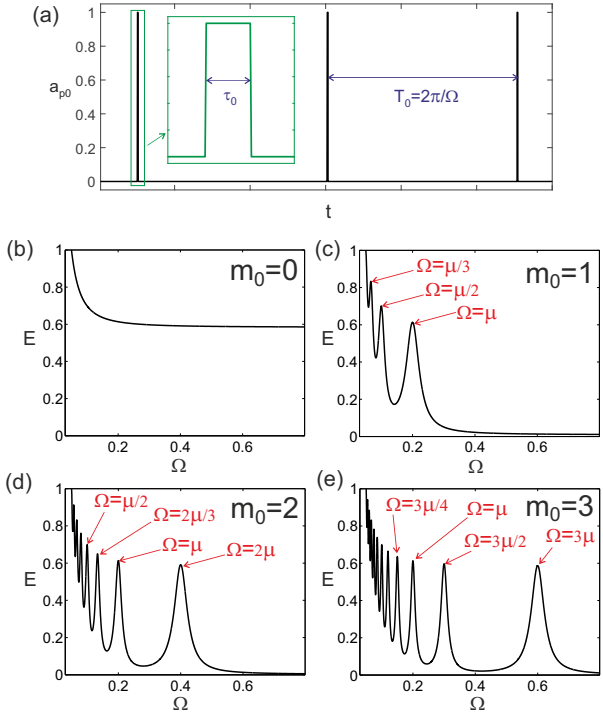}
\caption{(Color online) The driving force amplitude $a_{p0}(t)$ as a function of time is shown on panel (a). The dependencies of the average energy in the system defined as $E=\sum_n \frac{1}{T_0}\int_{0}^{T_0} |u_n|^2 dt $ are shown in panels (b)-(e) for $m=0, 1, 2, 3$ correspondingly. The parameters of the system and the driving force are $\gamma=0.025$, $\mu=0.2$, $\sigma=5$, $\alpha=0$, $\tau_0=0.5$.     } \label{fig12}
\end{figure}

To check the analytics we compared the analytical results against direct numerical simulations. In the numerical the driving force $a_n=A\xi(n)\exp(-i\omega_p t)$ is taken as a function of $\Omega$ $A= \frac{A_0}{\Omega} \sum_l  \Theta(t-\frac{2\pi l}{\Omega})-\Theta(t-\frac{2\pi l}{\Omega}-\tau_0  )  $ where $\tau_0$ is the duration of the individual pulses in the train, $\Theta$ is the Heaviside step function, $A_0$ is a constant. This choice of the pump provides that the spectral intensity  of the pump $|{\cal A}_0|^2$ at low frequencies does not depend on $\Omega$. The temporal dependency of the pump amplitude $A$ is illustrated in Fig.~\ref{fig12}(a). The spatial distribution of the pump is taken in the form of Wannier functions $\xi=W_{m0}(n)$, the pump frequency is taken $\omega_p=2\sigma$. 

The results of the numerical simulations for $\tau=0.5$ are shown in Fig.~\ref{fig12}(b)-(e) for different positions on the pump ($m_0=0, \, 1, \, 2, \, 3$). The parameter $A_0$ is chosen to provide that for $\Omega=0.05$ the average field intensity is $E=1$. One can see that, indeed, for $m_0=0$ the dependency $E(\Omega)$ does not show any resonances whereas for $m_0=1$ the resonances take place at $\Omega=\frac{\mu}{l}$, for $m_0=2$ -- at $\Omega=\frac{2 \mu}{l}$ and so on. Let us remark that the increase of $E$ at low frequencies is associated with the increase of the power of the pump (the intensity of the pump averaged over a period scales as $1/\Omega$).

\begin{figure}
\includegraphics[width=0.45\textwidth]{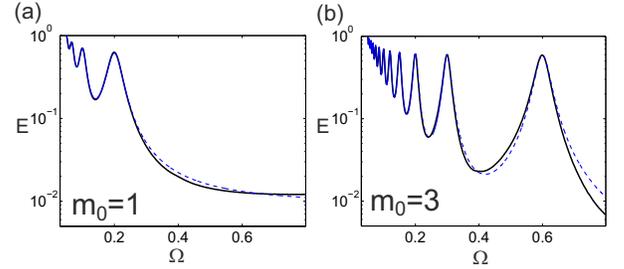}
\caption{(Color online) Panel (a) and (b) shows the same as Fig.~\ref{fig12}(c) and Fig.~\ref{fig12}(e) but for the duration of the individual pulses $\tau_0=5$. The solid dark lines are the numerics and thinner dashed blue lines are the dependencies plotted by (\ref{eq_app_7}). The scale of the vertical axis is logarithmic.    
     } \label{fig13}
\end{figure}

These numerical dependencies $E(\Omega)$ were checked against the dependencies given by (\ref{eq_app_7}) and for this choice of the parameters the agreement is excellent. For larger duration of the individual pulses $\tau_0$ the difference between the numerics and the dependencies calculated by (\ref{eq_app_7}) becomes noticeable, see Fig.~\ref{fig13} showing the case of $m_0=1$ and $m_0=3$. 

Let us remark that for general form of the pump spatial distribution all Wannier modes are excited (of course with different efficiency) and thus in this case we can expect the overlap of resonances of the pump with the first, second, third and all the rest Wannier-Stark states. This results in the resonances at the frequencies $\Omega=\frac{m \mu}{l}$. This case is discussed in Section II.

\end{document}